%% file: eprint.tex
\documentclass[10pt, paper=a4, UKenglish]{article}
\usepackage{graphicx}
\def\Title#1{\begin{center} {\Large #1 } \end{center}}
\def\Author#1{\begin{center}{ \sc #1} \end{center}}
\def\Address#1{\begin{center}{ \it #1} \end{center}}

\newcommand\pubblock{\rightline{\begin{tabular}{l} Proceedings of CTD 2020\\ \pubnumber\\
         \pubdate  \end{tabular}}}

\newenvironment{Abstract}{\begin{quotation} \begin{center} 
             \large ABSTRACT \end{center}\bigskip 
      \begin{center}\begin{large}}{\end{large}\end{center} \end{quotation}}

\newenvironment{Presented}{\begin{quotation} \begin{center} 
             PRESENTED AT\end{center}\bigskip 
      \begin{center}\begin{large}}{\end{large}\end{center} \end{quotation}}

\def\Acknowledgements{\bigskip  \bigskip \begin{center} \begin{large}
      \bf ACKNOWLEDGEMENTS \end{large}\end{center}}

\input econfmacros.tex

\usepackage{color}
\usepackage{lineno}
\usepackage{subfig}
\usepackage{hyperref}
\usepackage{graphbox}

\newcommand\pubnumber{PROC-CTD2020-34}

\newcommand\pubdate{\today}

\def\affiliation{ 
        $^1$Middle East Technical University, Ankara, Turkey \\
        $^2$STB Research, Ankara, Turkey \\
        $^3$CERN, Geneva, Switzerland \\
        $^4$gluoNNet, Geneva, Switzerland \\
        $^5$DESY, Hamburg, Germany \\
        $^6$Lancaster University, Lancaster, UK \\
        $^7$University of Padua, Padua, Italy \\
        $^8$California Institute of Technology, Pasadena, California, USA }

\newcommand{\conference}{Connecting the Dots Workshop (CTD 2020)\\
April 20-30, 2020}

\usepackage{fancyhdr}
\definecolor{mygrey}{RGB}{105,105,105}
\begin{document}


\large
\begin{titlepage}
\pubblock

\vfill
\Title{A Quantum Graph Neural Network Approach \\ 
to Particle Track Reconstruction
}
\vfill

\Author{Cenk T\"{u}ys\"{u}z$^{1,2}$, Federico Carminati$^{3}$, Bilge Demirk\"{o}z$^{1}$, Daniel Dobos$^{4,6}$, Fabio Fracas$^{3,7}$, Kristiane Novotny$^{4}$, Karolos Potamianos$^{4,5}$, Sofia Vallecorsa$^{3}$, Jean-Roch Vlimant$^{8}$}
\Address{\affiliation}
\vfill

\begin{Abstract}
Unprecedented increase of complexity and scale of data is expected in computation necessary for the tracking detectors of the High Luminosity Large Hadron Collider (HL-LHC) experiments. While currently used Kalman filter based algorithms are reaching their limits in terms of ambiguities from increasing number of simultaneous collisions, occupancy, and scalability (worse than quadratic), a variety of machine learning approaches to particle track reconstruction are explored. It has been demonstrated previously by HEP.TrkX using TrackML datasets, that graph neural networks, by processing events as a graph connecting track measurements can provide a promising solution by reducing the combinatorial background to a manageable amount and are scaling to a computationally reasonable size. In previous work, we have shown a first attempt of Quantum Computing to Graph Neural Networks for track reconstruction of particles. We aim to leverage the capability of quantum computing to evaluate a very large number of states simultaneously and thus to effectively search a large parameter space. As the next step in this paper, we present an improved model with an iterative approach to overcome the low accuracy convergence of the initial oversimplified Tree Tensor Network (TTN) model.
\end{Abstract}

\vfill

\begin{Presented}
\conference
\end{Presented}
\vfill
\end{titlepage}
\def\thefootnote{\fnsymbol{footnote}}
\setcounter{footnote}{0}
%

\normalsize 



\section{Introduction}
\label{intro}

Particle track reconstruction is a common problem in most particle physics experiments. At collider experiments, particles are accelerated to speeds very close the speed of light and collide in bunches at rates of $\mathcal{O}(10 MHz)$. In each collision, new particles are created and they scatter in all directions. These particles pass through particle tracking detectors and create signals which are called \textit{hits}. Particle tracking algorithms aims to distinguish these signals and identify the trajectory of the particles.

CERN created the kaggle TrackML challenge in 2018 to invite researchers from all backgrounds to solve the particle track reconstruction problem \cite{trackml}.  Later, the simulated dataset created for the challenge became popular among researchers to test and benchmark new ideas in the field.

In a few years, the Large Hadron Collider (LHC) at CERN will be upgraded to become the High Luminosity Large Hadron Collider (HL-LHC). This upgrade, which will ramp up the rate of collisions, comes with many challenges, one of which is the particle track reconstruction problem \cite{ref-hilumi}. Although, the novel particle tracking algorithms can manage the current rate of collision rates, they suffer from higher collision rates as they scale worse than polynomially. Therefore, the search for faster particle track reconstruction algorithms is very important.

There are many initiatives to bring faster solutions to particle track reconstruction. Researchers in the field explore novel methods such as Machine Learning and Quantum Computing. The HepTrkX team proposed a Graph Neural Network (GNN) approach to solve the particle track reconstruction problem using the kaggle TrackML challenge dataset \cite{heptrkx}. Other researchers also proposed new methods using Quantum Annealing to tackle the challenge \cite{qalg3}.

In this work, we present our updated results on the Quantum Graph Neural Network approach, which combines the novel GNN method of the HepTrkX project with the quantum circuit model \cite{heptrkx-quantum}.


\section{The Dataset and Classical Approach}
\label{dataset-classical}

This work uses the publicly available TrackML dataset \cite{trackml}. The dataset contains spatial coordinates of each hit created by particles which are created during simulated collisions. Each of these collisions are called as events and the dataset contains 10000 event files. Among these files only 100 events are used due to restrictions in simulation times. Quantum Circuit simulations on both CPU and GPU use extensive resources and time as the number of qubits increase. In the case of our model, it takes around a week of CPU time to train the model over 100 events for a single epoch.

\begin{figure}[!htb]
  \centering
  \includegraphics[width=0.62\linewidth]{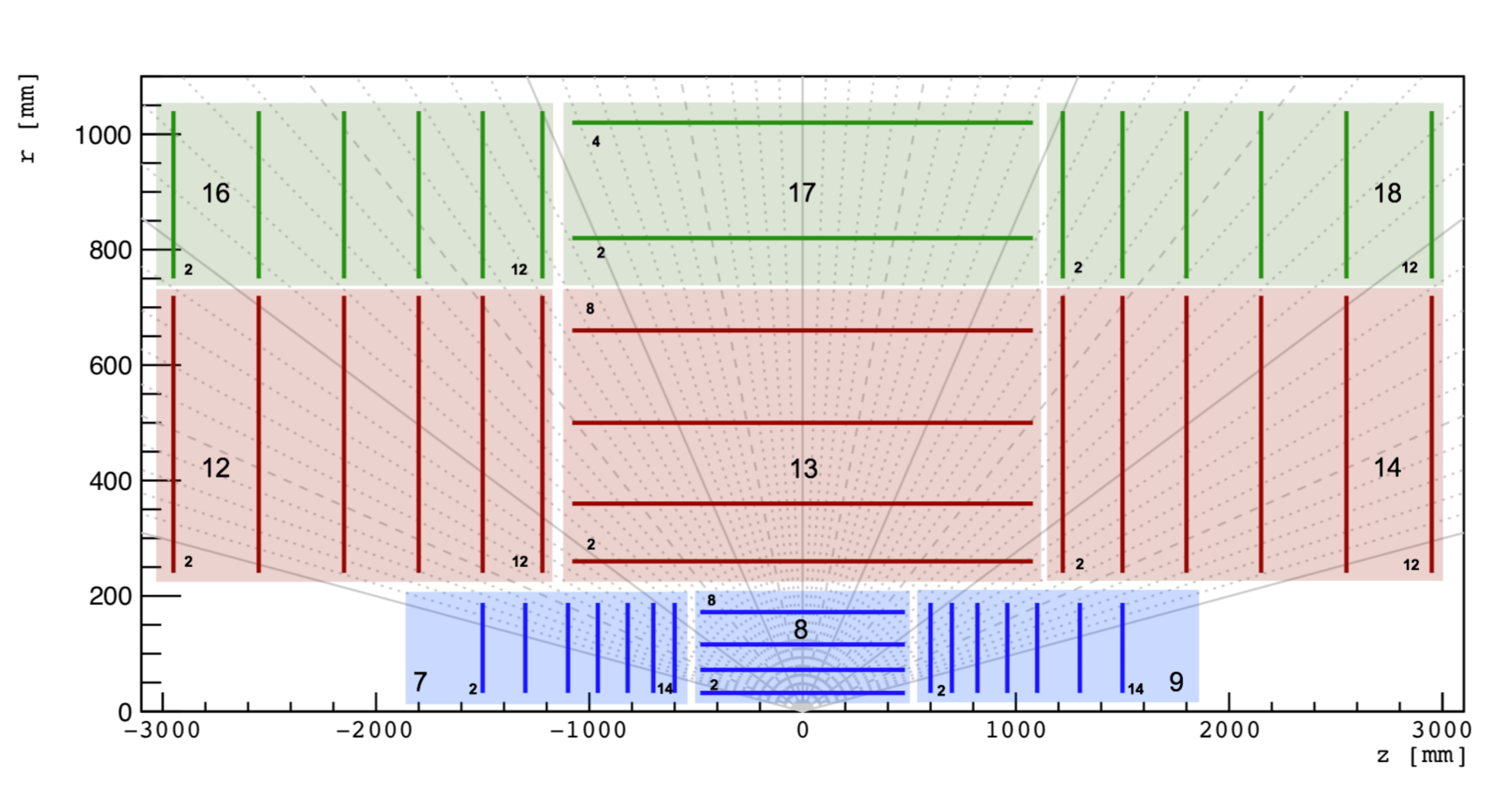}
  \caption{TrackML Detector Layout \cite{trackml}. }
  \label{fig:trackml-detector}
\end{figure}

\noindent The TrackML data is created using a simulated detector having a similar geometry to most LHC experiments. The detector has horizontal (barrel) layers near the center of collisions and vertical (endcap) layers outside. The particle beams propagate along the $z$-axis and collide around $z=0$. The TrackML detector layout can be seen in Figure~\ref{fig:trackml-detector}. The produced particles of these collisions scatter through all directions. This work only uses the barrel region hits to simplify the track reconstruction problem as the ambiguity of the tracks is much higher for endcap hits.

\noindent The HepTrkX GNN approach converts the dataset consisting of spatial coordinates of hits to a graph dataset. A set of loose selection criteria is applied to each hit in order to eliminate illogical connections in the graph. This is way, it is ensured that the preprocessing step takes a very short amount of time and graphs are not fully connected in order to minimize run time. The selection criteria determined by the HepTrkX team is respected in this work and can be seen in Table~\ref{tab:table1} \cite{heptrkx}.

\begin{table}[!htb]
  \begin{center}
      \caption{Selection applied to TrackML dataset for preprocessing.}
    \begin{tabular}{|c|c|}
\hline
$\left|p_T\right|$ & $>1 GeV$     \\\hline
$\Delta\phi$ & $<0.0006$  \\\hline
$z_0$ & $<100 mm$  \\\hline
$\eta$ & $[-5,5]$  \\\hline
    \end{tabular}
    \label{tab:table1}
  \end{center}
\end{table}

\noindent The coordinate definitions used to describe the data is as follows. $\phi$ is the angle along the transverse plane ($xy$ plane) and $\left|p_T\right|$ is the magnitude of momentum along the same plane. $\eta$ is the psuedorapidity measures the azimuthal angle with respect to the beam axis (z-axis). 

\noindent An event contains $\sim 8k$ hits, therefore the model requires a huge amounts of memory to load a single event. As the detector geometry is cylindrically symmetric along the $\phi$ and the $z$ direction, the data set is divided into 8 in the $\phi$ and into 2 in the $z$ direction to reduce the size of each event. Therefore, the new graph dataset contains 1600 subgraphs of originated from 100 events. 

\begin{figure}[!htb]
  \centering
  \includegraphics[width=\linewidth]{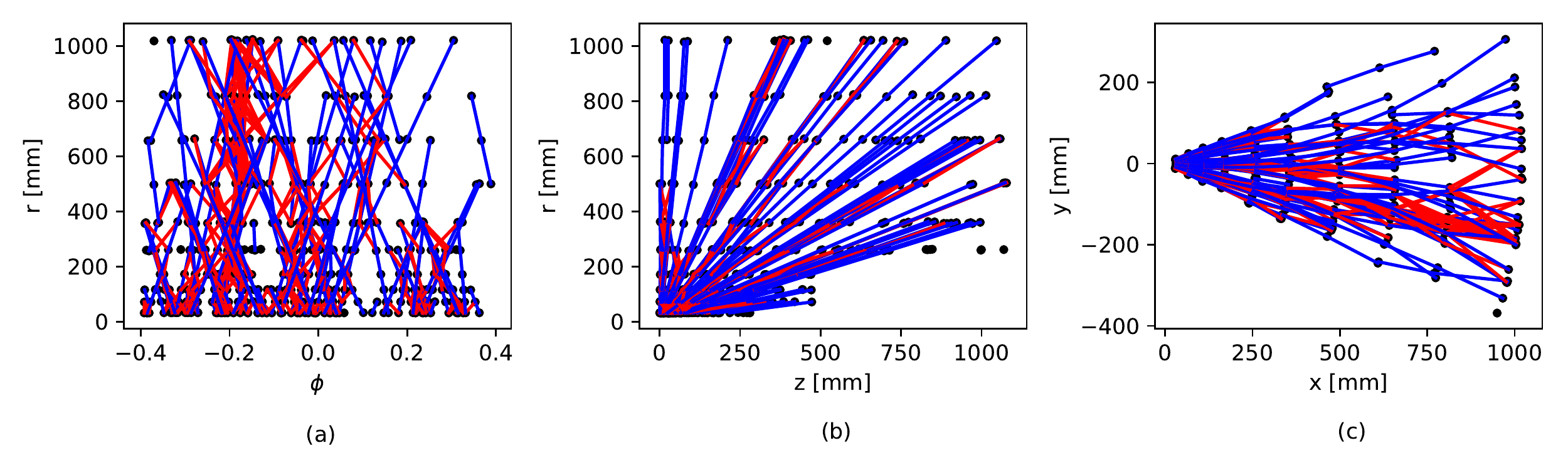}
  \caption{1 of 16 subgraphs created from a single event. (a,b) are subgraphs in cylindrical coordinates and (c) is a subgraph in Cartesian coordinates. Red represents Ground Truth, while Blue shows Fake edges created using loose cuts.}
  \label{fig:graph}
\end{figure}

\begin{figure}[!htb]
  \centering
  \includegraphics[width=\linewidth]{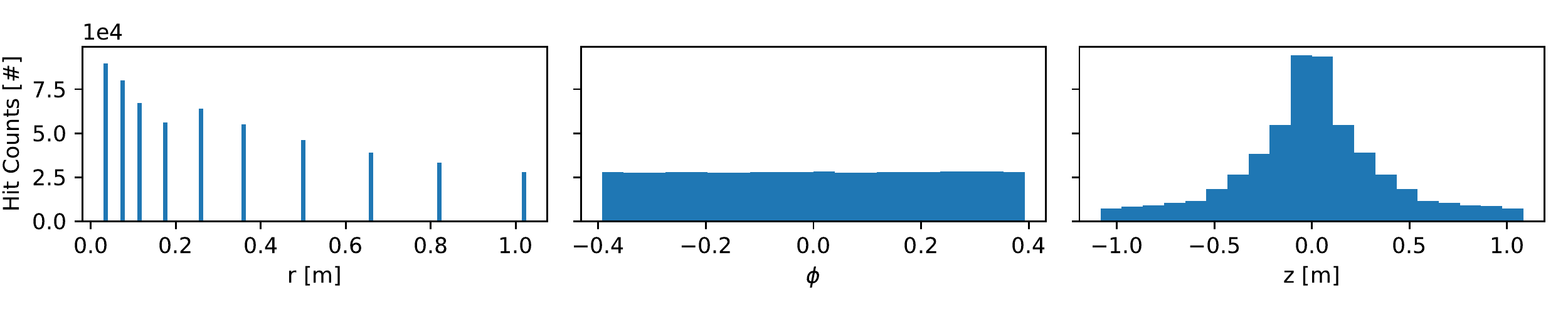}
  \caption{Histogram of hits crated using 1600 subgraphs in cylindrical coordinates.}
  \label{fig:data}
\end{figure}

\noindent An example subgraph can be seen in Figure~\ref{fig:graph} and the distribution of the subgraphs for each coordinate variable, in Figure~\ref{fig:data}. The distribution of r and z can easily be explained by referring to the geometry of the detector in Figure~\ref{fig:trackml-detector}. The distribution in $\phi$ is uniform as expected since the geometry is symmetric along the transverse plane.
 
\section{The Quantum Graph Neural Network Approach}
\label{qgnn}

Quantum circuits have been previously shown to be able to handle classification tasks previously \cite{mps,ttn}. Although, Quantum Machine Learning has not yet been shown to outperform classical Machine Learning, scientist are trying new methods achieve speed-ups for certain tasks. High Energy Physics is no exception to this trend \cite{hep-qml}. In this work, we explore the use of Quantum Circuits to perform track segmentation.

\begin{figure}[!htb]
\centering
\includegraphics[width=\linewidth]{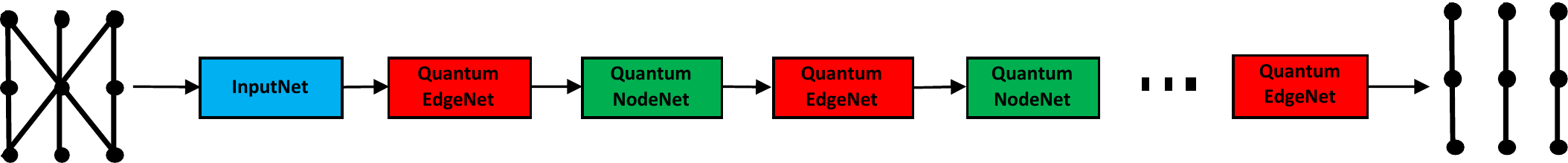} \\
\caption{The Quantum Graph Neural Network model used in this work.} 
\label{fig:network}       
\end{figure}

\noindent In order to integrate Quantum Computing and GNNs, the Neural Networks of Edge and Node Network have been repleaced by two Quantum Circuits. The new model flow can be seen in Figure~\ref{fig:network}. There are many Quantum Circuits that perform binary classification tasks in the literature \cite{mps,ttn}. The Tree Tensor Network (TTN) model is chosen due to its simplicity among these circuits. The TTN circuits are implemented using Pennylane along with its Tensorflow interface. Pennylane provides the necessary gradients of the Quantum Circuits during the training step while Tensorflow is used for optimization and constructing the model pipeline \cite{tensorflow, pennylane}.

\noindent The new hybrid model, which is called the Quantum Graph Neural Network (QGNN) takes a graph as an input. The model begins with an Input Layer, which is a single layer Neural Network used to increase the dimension of the input and increases the dimension of the data to the required number from 3. 3 is the original size of the input as there are 3 spatial coordinates for each hit. Then a Quantum Edge Network (QEN) is applied to each edge of the input graph which outputs an edge feature. The output edge feature is passed into each edge to update their value. This information is used by the Quantum Node Network (QNoN), which is applied to each node of the graph. The output of the QNoN is used to update nodes in the hidden layers. Recursive iterations of Edge and Node Networks allow the information to propagate from lower detector layers to above layers. Finally, a QEN is applied to obtain the final segment classification.

\noindent The use of TTN inside QEN and QNoN is almost identical except the amount of qubits used. QEN is applied to each edge one by one, therefore the size of input is 6 (amount of spatial coordinates for 2 nodes) in the case of no hidden dimensions. These coordinates are mapped to $[0,2\pi]$. Then, the $R_y(\theta)$ gate is used encode the information to each qubit. The encoding can be represented with the simple equation below.

\begin{equation}
\label{eq1}
R_y(\theta)\ket{0} = \cos(\theta/2)\ket{0} + \sin(\theta/2)\ket{1}
\end{equation} 

\begin{figure}[!htb]
  \centering
  \subfloat{\includegraphics[align=c,width=0.47\linewidth]{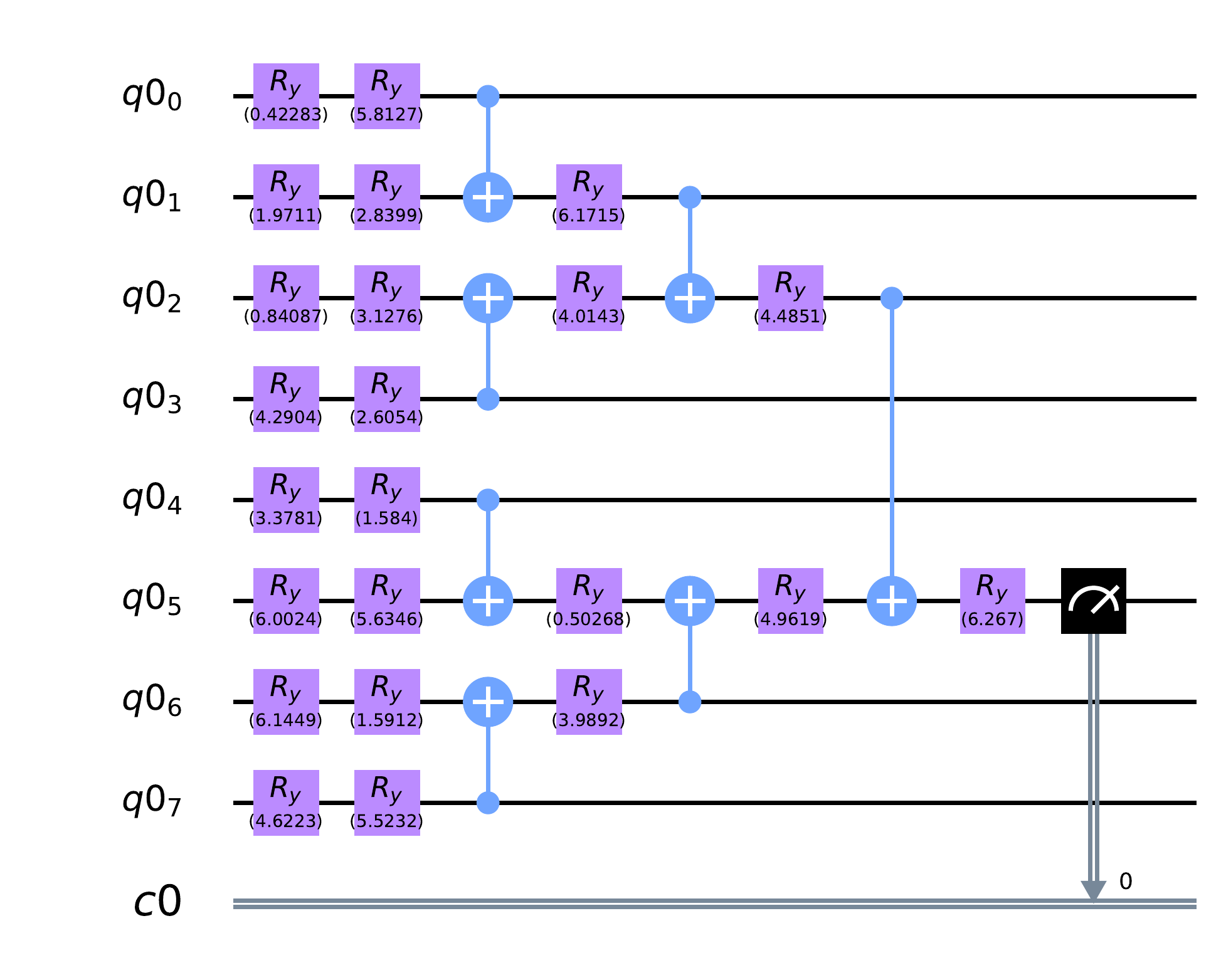}}
 \qquad
  \subfloat{\includegraphics[align=c,width=0.4\linewidth]{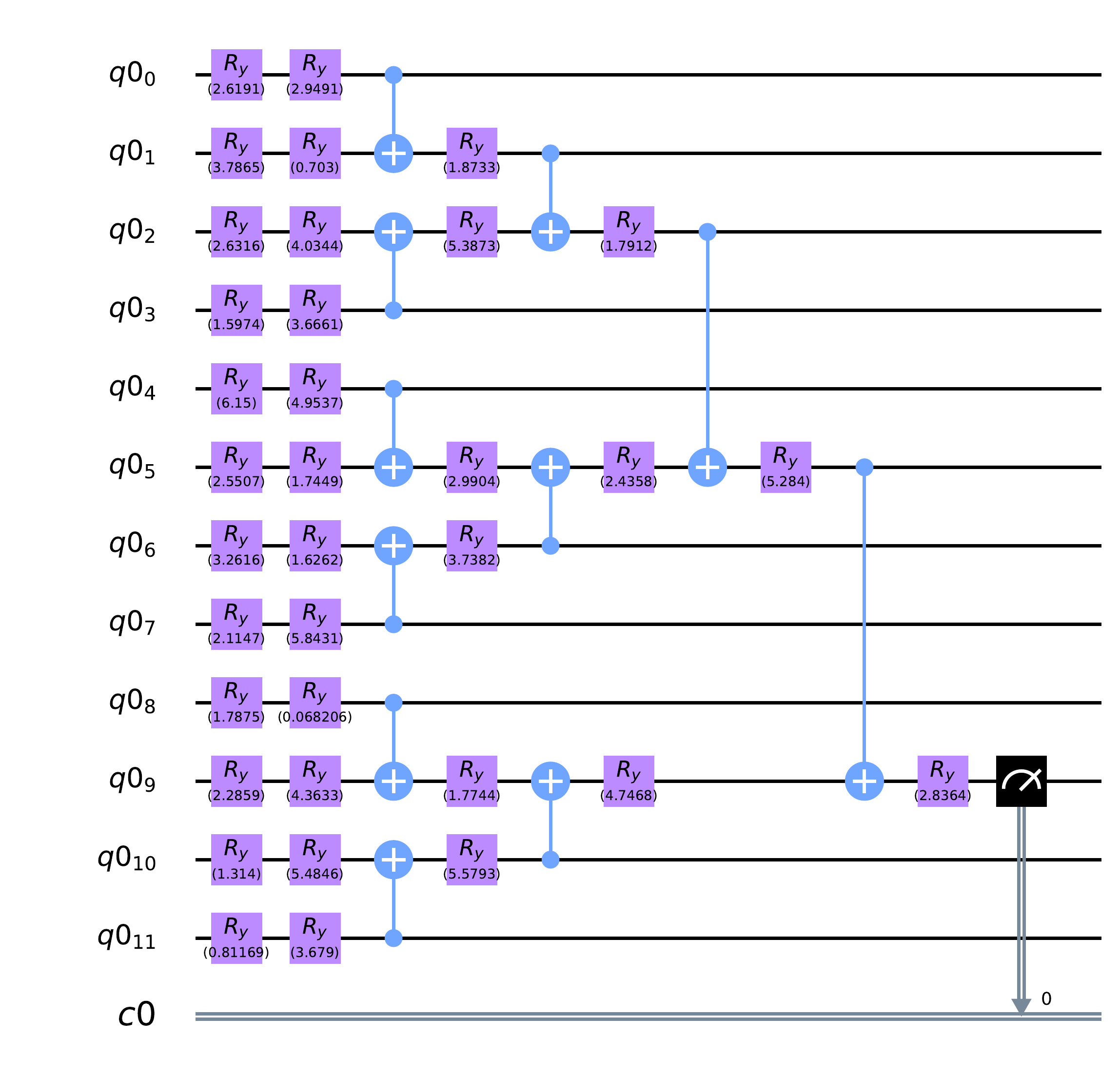}}
       \caption{Quantum Circuit of QEN is given on the left. Quantum Circuit of QNoN is given on the right. The numerical values are from an example data.}
  \label{fig:circuit}
\end{figure}

\noindent The TTN circuit is applied afterwards and a measurement is taken. To perform a simple state tomography on the output qubit, the process is repeated many times and the expectation value is calculated. This value is used as the edge feature. By applying the QEN to each edge, the edge features for all edges are calculated. The same process is repeated for the QNoN case. The only difference between the QEN and the QNoN is the size of the circuit and the inputs being nodes, rather than the edges. At the final step of the QGNN, edge features for all initial edges are obtained. These values are used to calculate a weighted binary cross entropy loss taking into account the ratio of the true edges to the false edges. Then the ADAM optimizer of the Tensorflow is used to update the variables of the TTN circuits of both QEN and QNoN. In this work, the QGNN model with the Quantum Circuits of QEN and QNoN used shown in Figure~\ref{fig:circuit} is used with one hidden dimension.

\begin{figure}[!htb]
  \centering
  \subfloat[Validation Loss]{\includegraphics[width=0.47\linewidth]{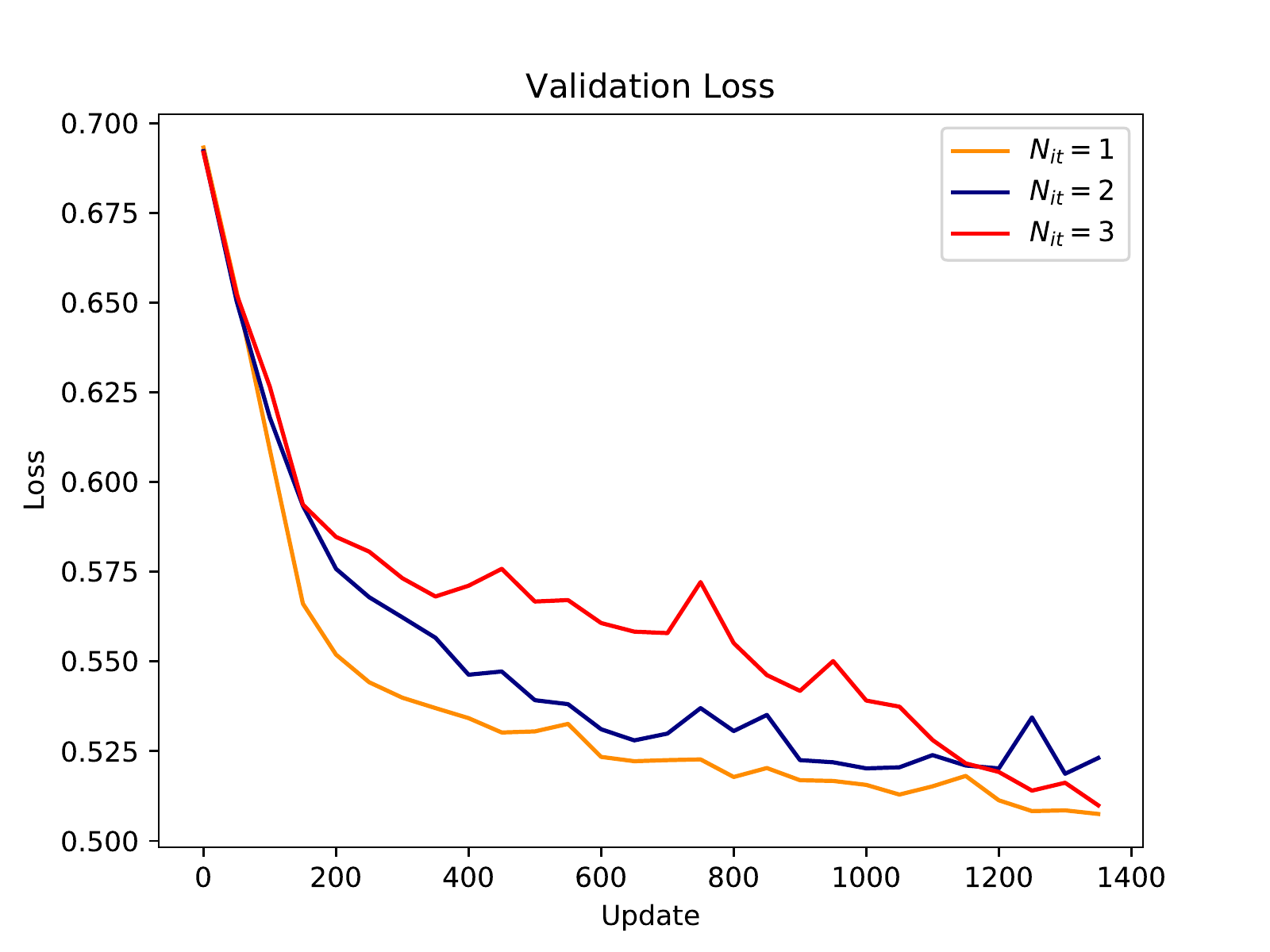}}
 \qquad
  \subfloat[Validation AUC]{\includegraphics[width=0.47\linewidth]{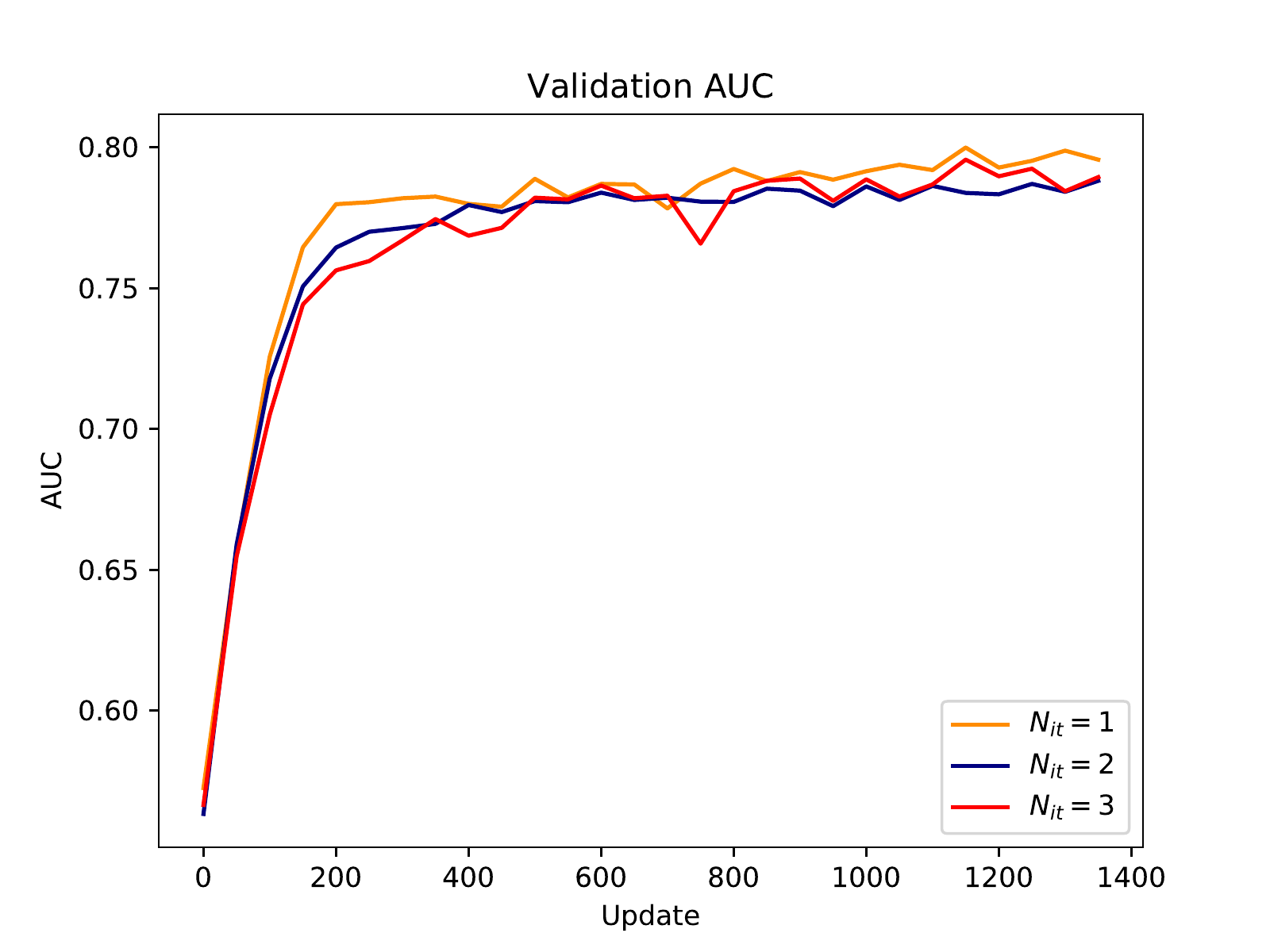}}

       \caption{Validation Set results for a single epoch of training with different iteration values.}
  \label{fig:results}
\end{figure}

\noindent 1400 subgraphs are used to train the QGNN and 200 subgraphs are used in the validation set for a single epoch. 3 independent experiments were conducted to test different iterations. The training results are shown in Figure~\ref{fig:results}.

\noindent Figure~\ref{fig:results} shows that the model can achieve an Area under the ROC curve (AUC) of 0.80 and a binary cross entropy loss of 0.5. While 1.0 is the perfect score for AUC, the model seems to perform well considering its simplicity. It was expected that the model performance to get better as the number of iterations increases. However, higher iteration runs performed similar to $N_{it}=1$ or even worse. There are two main reasons leading to this. First, simple TTN models do not represent the data more than the current best results, therefore it defines a less than perfect ending point for the training. The second issue is related to the vanishing gradient problem. It is known that Recurrent Neural Networks suffer from this problem and the QGNN is no exception. Therefore, as the $N_{it}$ increases the learning rate slows down. These two major issues will be investigated deeply in future work.
 
\section{Future Work}
\label{fw}
The work presented here is the first full demonstration of a Quantum Graph Neural Network that can classify track segments with good precision and accuracy. There is a need for more detailed work to further increase the performance of the model. Future work should include the following;
\begin{itemize}
    \item Extending the size of the hidden dimension, which was limited to 1 in this work.
    \item A detailed analysis of the vanishing gradient problem.
    \item Trying different Quantum Circuit architectures to achieve better accuracy.
\end{itemize}


\section{Conclusions}

This work presents a first implementation of a Quantum Graph Neural Network dedicated for solving track reconstruction problem. Although the presented model uses the simplest form of Quantum Circuits in literature and the minimum number of qubits that can be used, it performs well. Current results show that a simple QGNN model performs very similar to sophisticated track reconstruction models. However, there is still more to be done to achieve this. It is also important to note that this work currently only uses simulations and does not consider practical applicability or possible speed-ups.


\Acknowledgements
Part of this work was conducted at "\textit{iBanks}", the AI GPU cluster at Caltech. We acknowledge NVIDIA, SuperMicro and the Kavli Foundation for their support of "\textit{iBanks}". This work was partially supported by Turkish Atomic Energy Authority (TAEK) (Grant No: 2017TAEKCERN-A5.H6.F2.15).



\end{document}

%% file: econfmacros.tex
\def\beq{\begin{equation}}
\def\eeq#1{\label{#1}\end{equation}}
\def\eeqn{\end{equation}}

\def\beqa{\begin{eqnarray}}
\def\eeqa#1{\label{#1}\end{eqnarray}}
\def\eeqan{\end{eqnarray}}

\let\bar=\overbar

\def\ket#1{\left| {#1} \right\rangle}

\def\Dslash{\not{\hbox{\kern-4pt $D$}}}
\def\dslash{\not{\hbox{\kern-2pt $\del$}}}

\def\msb{{\bar{\ssstyle M \kern -1pt S}}}

